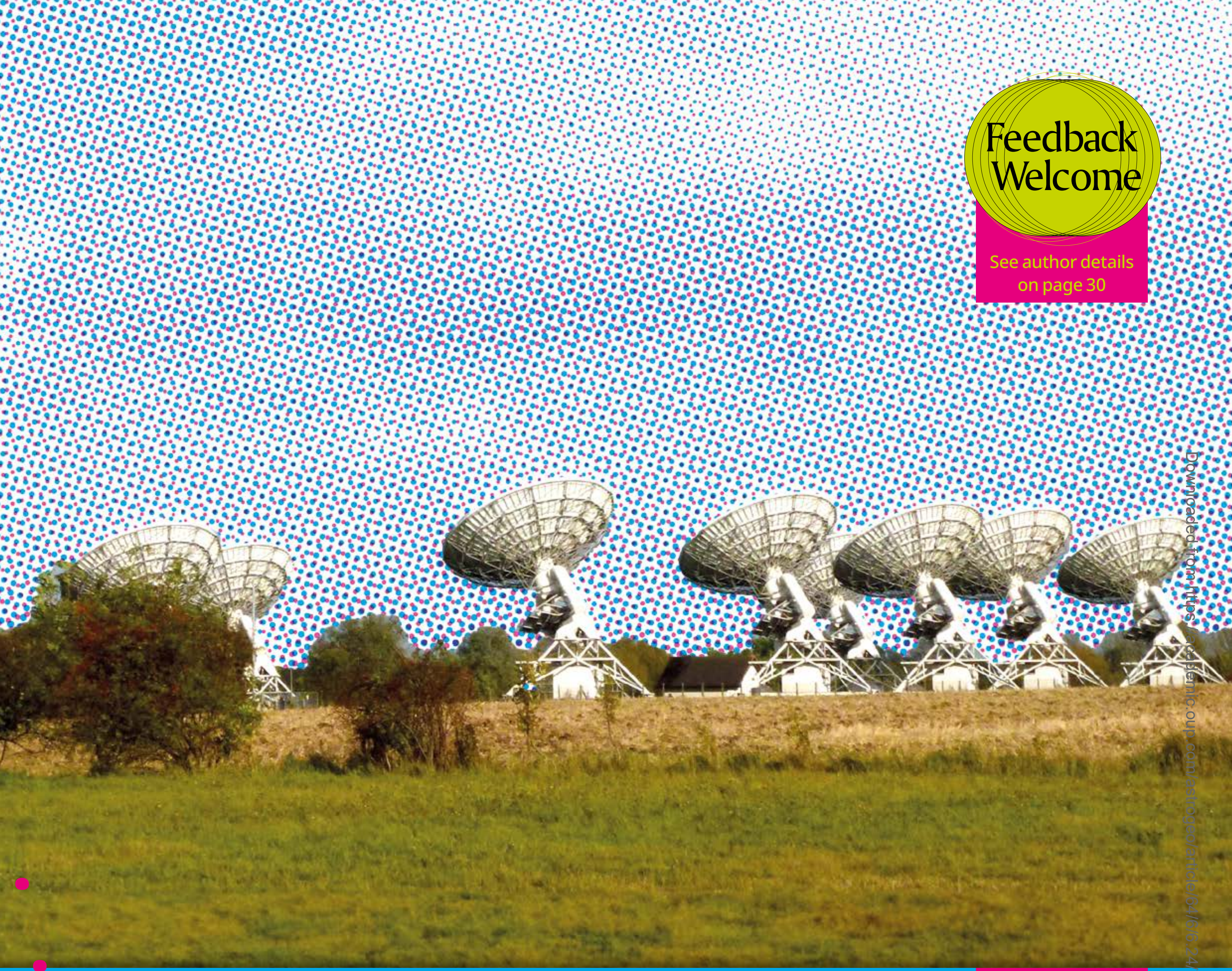

Feedback Welcome

See author details on page 30

Rob Fender, Assaf Horesh, Phil Charles, Patrick Woudt, James Miller-Jones and Joe Bright make the case for a dedicated radio transients monitoring array in the southern hemisphere

# Filling the radio transients gap

Extreme astrophysical events produce transient radio emission, from their extremes of temperature and from the acceleration of particles to relativistic energies. The next generation of large array telescopes such as the Square Kilometre Array (SKA) will not have the programme time to comprehensively monitor such events, but there is a case for a small radio telescope array in the southern hemisphere with operations dedicated to rapid follow-up and monitoring of astrophysical transients. The science harvest from such a facility would be very large, as we can judge by using the Arcminute Microkelvin Imager – Large Array (AMI-LA), as an outstanding example of how such a programme is already being operated in the north.

1 *The AMI-LA radio array at the Mullard Radio Astronomy Observatory, Cambridge, UK. See also aerial photo in 'A comparison of three small radio arrays'* (cmglee; CC BY-SA 3.0).

As a concept and discussion piece, the authors welcome comments and feedback on the ideas in this article.

Extreme astrophysical events – such as relativistic flows, cataclysmic explosions, black hole accretion and gravitational wave bursts from merging neutron stars – are frontier astrophysics in the 21st century. The extremes of physics – density, temperature, pressure, velocity, gravitational and magnetic fields – experienced in these environments are beyond anything achievable in any laboratory on Earth, and they provide a unique glimpse at the laws of physics operating in extraordinary regimes. Simply put, transient astrophysical phenomena signpost the most extreme physics in our universe since the Big Bang.

All of these events are associated with transient radio emission, a tracer both of the acceleration of particles to relativistic (TeV) energies, and coherent emitting regions with effective temperatures in excess of $10^{35}$K. By studying radio bursts from these phenomena we can pinpoint the sources of explosive events, understand the budget of kinetic feedback by such events in the ambient medium, and potentially probe the physical state of the universe back to the epoch of reionisation, less than a billion years after the Big Bang.

Some of the most important results in high-energy astrophysics in the 20th century arose from studies of radio emission from variable and transient sources. These results include the discovery of solar radio bursts (Appleton 1945), that of pulsars (Hewish *et al.* 1968), that of superluminal motions from both extragalactic and galactic relativistic jet sources (Rees 1966; Mirabel & Rodriguez 1994) and many other important discoveries.

In the 21st century, the scope of radio transients studies has further diversified, with prominent examples including the discovery of Fast Radio Bursts (FRBs; Lorimer *et al.* 2007), radio emission from Tidal Disruption Events (TDEs; Zauderer *et al.* 2011) and a long-lived radio afterglow from the gravitational wave LIGO/VIRGO transient GW170817, which resulted from the merger of two neutron stars (Hallinan *et al.* 2017). The astrophysics probed by the radio emission remains paramount – only radio observations can measure the total kinetic energy output of an event, measure the dispersion (and hence composition) of the intergalactic medium on cosmological scales, or directly resolve proper motions in relativistic outflows and hence the fundamental geometry of the transient. New avenues continue to open up, for example studies of radio flaring from low-mass dwarf stars, likely to be a signature of conditions inhospitable for any complex life on orbiting exoplanets.

## Radio astronomy in the 21st century: a road to larger facilities

Globally, the power and survey capability of radio astronomy is dominated already, as it will be in the SKA era, by a small number of very powerful facilities. These are, notably, the VLA, LOFAR, GMRT in the northern hemisphere and MeerKAT, ASKAP and ATCA in the south. Facilities with more niche operations reap huge harvests in narrower scientific fields (e.g. the 1000s of FRBs discovered with CHIME, designed primarily for HI work). Large facilities like the VLA and MeerKAT are able to perform cutting-edge science across the whole spectrum of astrophysical research, from cosmological targets to solar system objects. A major use of such facilities is to undertake wide and deep surveys of the extragalactic sky in order to understand the structure and evolution of our universe. This fantastically wide scientific capability means that these facilities are heavily oversubscribed from essentially all parts of the astronomical community – a sure sign of success.

The most commonly used metric when assessing the comparative performance of radio telescopes is the survey speed figure of merit (SSFOM, also known simply as 'survey speed'), which is the product of the sensitivity squared and the effective field of view (see e.g. Braun *et al.* 2019). In recent decades the drive to maximise the SSFOM has driven designs towards large numbers of relatively small dishes (~15m in the case of SKA1-MID), which provide both large fields of view (small dishes) and high sensitivity (for many of them), at the cost of increased data transport and computational load on the correlator (number of correlated baselines = N(N-1)/2 where N is number of antennas, so approximately $N^2$ for large N). This has led to extremely powerful and flexible facilities such as MeerKAT (64 13.5-metre dishes) and ASKAP (36 12-metre dishes, each equipped with a phased array feed) and, within this decade, the SKA (~200 13.5–15-metre dishes). Other arrays propose to take the concept even further, with the Deep Synoptic Array (DSA-2000) proposing to deploy 2000 five-metre dishes.

*"We propose a relatively small, low-cost imaging array dedicated to astrophysical transient follow-up and monitoring at radio wavelengths"*

## The divergence between radio and optical astronomy

Radio astronomy's use of interferometers means its path has diverged from that of, say, optical astronomy (parallels can also be seen at other wavelengths, e.g. X-ray). The most notable difference is that the development of increasingly large single optical telescopes (ESO's Extremely Large Telescope will have a 39-metre mirror and is due for completion around 2030) has happened alongside a proliferation of new small and medium-sized facilities. Many of the smaller facilities are robotic and are designed for a combination of rapid-response and/or high-cadence observations of variable sources. The discovery yield of optical transients, from supernovae to exotica such as neutron star mergers, has been largely driven by these facilities (examples include ATLAS, ZTF, ASASSN, PanSTARRS etc).

The reason for this divergence in paths is multifold, and part of the reason is that it takes some significant collecting area to make radio observations of interest, whereas very small (e.g. 40cm) optical telescopes can still detect optical transients. However, the central point is that, as they maintain the large fields of view while simultaneously being highly sensitive, a single large radio interferometer can, in principle, do everything. A large radio array can tackle this wide-field and high-cadence response to transients, while at the same time being capable of very deep stares. This is in stark contrast to optical astronomy, where very deep fields are necessarily observed by facilities with small fields of view (optical interferometers are technically much more challenging and not in widespread use). Furthermore, radio interferometers can sub-array in which multiple subsets of telescopes can image in different directions and/or different frequencies (with, of course, consequent reductions in sensitivity and image quality), which provides even more flexibility for observing astrophysical transients (whether multiple sources, or the same source simultaneously at different wavelengths). The lowest-frequency arrays (e.g. LOFAR, MWA, LWA, SKA1-LOW later this decade) are sensitive to an entire celestial hemisphere and with enough computing power the entire sky can, in principle, be imaged simultaneously (this is not yet achievable except with very coarse resolution – see e.g. the AARTFAAC project – due to the enormous computational expense of making large numbers of beams in software). Hence the flexibility of radio arrays greatly surpasses that of single large optical telescopes, and radio astronomy is in a good position.

## Why the cause for concern?

Our concern is that while future very large radio interferometers like SKA1-MID or ngVLA could potentially monitor large numbers of transients, due to the inherent flexibility of their designs, they will not actually do so because of the huge time pressures on their schedules combined with the large amount of time required to monitor transients to really

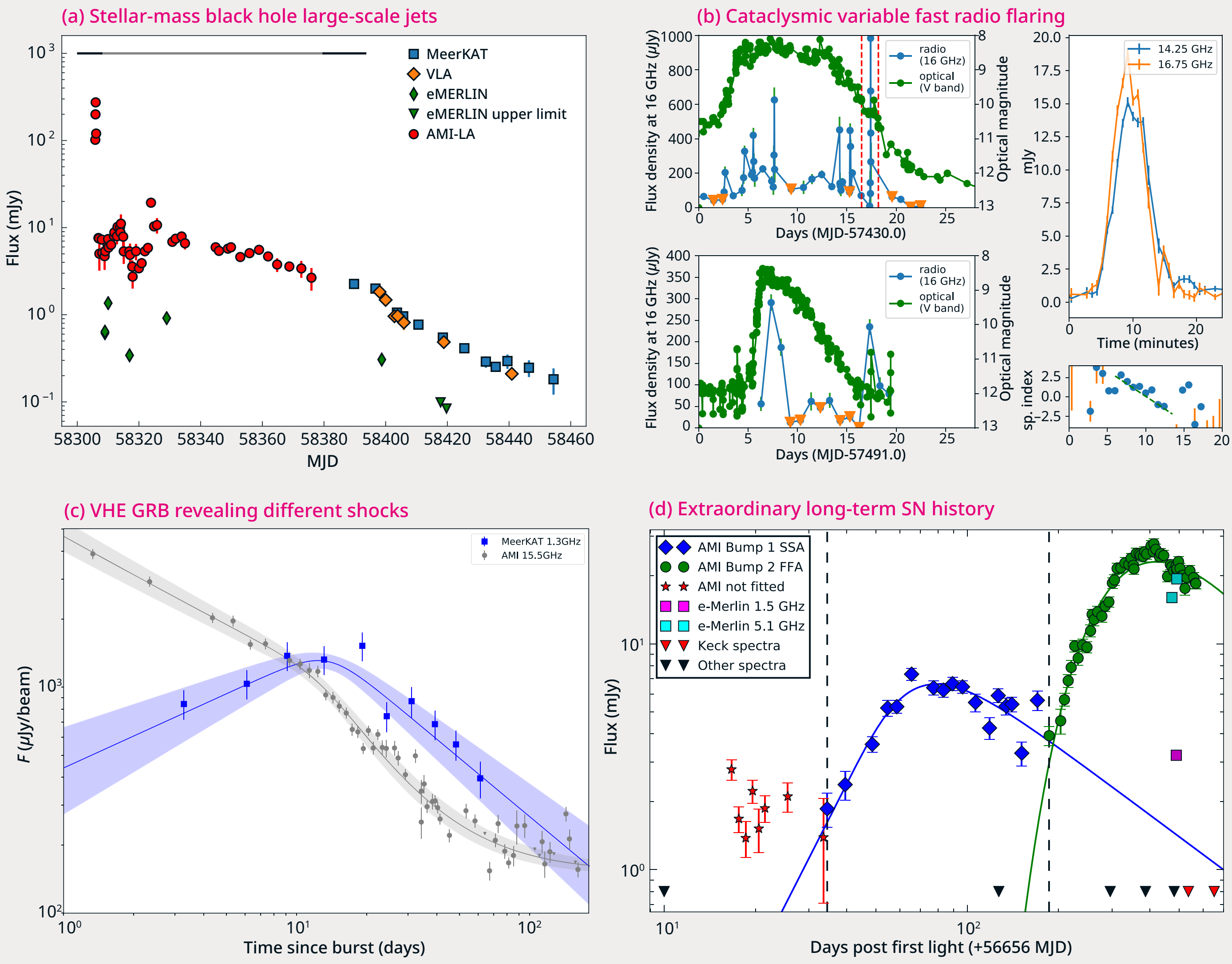

**2** *Examples of the extraordinary range of transient science done by AMI-LA, utilising its high cadence, rapid response and long-term monitoring.* **(a)** *Black hole MAXI J1820+070* (Bright et al. 2020), **(b)** *Cataclysmic variable SS Cyg* (Fender et al. 2019), **(c)** *Gamma Ray Burst GRB190829A* (Rhodes et al. 2020), *and* **(d)** *Supernova SN2014C* (Anderson et al. 2017).

extract the most valuable science. We suspect that very few users or time allocation committees will consider that extensive monitoring of the variability of a bright radio transient at a signal to noise of >>100 is a good use of the array's time, but neither will they be content to quasi-continuously give up 10 antennas from their project to a sub-array to perform this science. Astrophysical transients need rapid-response, high-cadence and long observations in order to catch the key moments that really provide the deepest insights into their nature. The motivation for this article is therefore to propose a relatively small, low-cost imaging array which is dedicated to astrophysical transient follow-up and monitoring at radio wavelengths.

## Huge success from a modest array: AMI-LA

Before we continue, let's look at an outstanding example of a small radio array, operating on a fraction of the budget of facilities like the VLA and MeerKAT, and yet which makes huge contributions to the field of radio transients. The array in question is the Arcminute Microkelvin Imager – Large Array, or AMI-LA, sited at and run by the University of Cambridge (figure 1).

This telescope has a long history, as the dishes were first constructed over 60 years ago, and have served in two different arrays with three different names: the Cambridge Five Kilometre Array, renamed to The Ryle Telescope (RT) and now AMI-LA. In the reconfiguration from the RT to AMI-LA the dishes were brought together on relatively short (maximum 110-metre) baselines, but were shifted from their original, purely east-west alignment, allowing better imaging at low astronomical declinations (see also aerial photograph in the box 'A comparison of three small radio arrays'). AMI-LA has eight dishes, a single set of single-polarisation wide bandwidth receivers with a central frequency of around 16GHz.

*"Not only is it the south where the SKA will operate, it also affords us a view of most of the galactic plane, the Magellanic Clouds and, crucially, the galactic centre"*

Prior to the transformation of the RT to AMI-LA, the transients work from the array was dominated by observations of highly variable radio-emitting X-ray binaries in our galaxy, in which transient radio emission was associated with the formation of relativistic jets. Some of these programmes persist to the AMI-LA (e.g. the very long-term monitoring of the archetypal black hole binary Cyg X-1 and archetypal 'microquasar' GRS 1915+105; Pooley & Fender 1997). These high-cadence long-term data sets from the RT were key to unlocking the connection between accretion states and the form and power of associated relativistic jet formation. The insights into black hole accretion gained from this programme (e.g. Fender *et al.* 2004) have been shown to be applicable to black holes on all mass scales from the most powerful AGN to Tidal Disruption Events [TDEs] (e.g. Marscher *et al.* 2002; Mummery 2021; Fernandez-Ontiveros & Muñoz-Darias 2021). These early results provided an insight into the power of long-term monitoring of transients. Beyond

X-ray binaries, AMI-LA has provided breakthrough observations of Cataclysmic Variables (e.g. Mooley *et al.* 2017), TDEs (e.g. Horesh *et al.* 2021; Sfaradi *et al.* 2022; Rhodes *et al.* 2023), Supernovae (Anderson *et al.* 2017), Fast Blue Optical Transients (Bright *et al.* 2022), nearby Flare Stars (Fender *et al.* 2015) and Gamma Ray Bursts (Rhodes *et al.* 2020; Bright *et al.* 2023) – see figure 2 for some snapshot examples.

In the case of TDEs, high-cadence observations of AT2022cmc (Rhodes *et al.* 2023; figure 3) indicated day timescale variability revealing a compact relativistic jet. Long-term monitoring combined with high-cadence observations helped uncover some of the first delayed radio flares in TDEs (Horesh *et al.* 2021; Sfaradi *et al.* 2022, 2023). The combination of long-term high-cadence observations also proved to be critical in identifying supernovae interacting with multiple CSM shells (observed as multiple bumps in the radio light curves), otherwise missed by sparse observations carried out by larger facilities (as in the case of SN2014C; Anderson *et al.* 2017). While multi-band sparse sensitive radio observations (provided by facilities such as the VLA and ATCA) are key for understanding the physics, continuous observations by AMI-LA can fill the gap, and provide essential information that is crucial for the interpretation (e.g. Sfaradi *et al.* 2022; figure 4).

The most recent GRB observations are worth highlighting. It was the rapid response and high-cadence observations possible with AMI-LA that provided the best ever coverage of a reverse shock from a bright, long-duration GRB (figure 5).

In the past decade, AMI-LA has spent some 20–50% of its observing time on astrophysical variables and transients. In many cases, it has made transient scientific breakthroughs which would have been inaccessible to its much larger and more powerful northern competitor, the VLA, simply due to the high-cadence and long-term monitoring it provides. The previous statement is of course highly subjective, but there are multiple examples of phenomena discovered by AMI-LA alone even while the target in question was a potential target for the VLA (i.e. was at an accessible declination). The main reason for this is the extreme flexibility in the scheduling, which allows for rapid response and very high-cadence observations when an interesting astrophysical source is active. AMI-LA also has the AMI Large Array Rapid Response Mode (ALARRM) in which it robotically responds to Swift X-ray triggers with no human intervention (note this mode is not always on). It was the first radio array to implement such a mode for transient response (Staley *et al.* 2013); inspired by this work, analogous automatic response modes were implemented on the Murchison Widefield Array (MWA; Hancock *et al.* 2019) and the Australia Telescope Compact Array (ATCA; Anderson *et al.* 2021). The AMI-LA Transients ADS catalogue for the past ten years (likely to be somewhat incomplete) can be found at tinyurl.com/AMITransients, and demonstrates the high impact of its science.

In the past year or so the reanimated Allen Telescope Array (ATA) has also begun flexible observations of radio transients on a shared-risks basis (and provided a major contribution to the GRB reverse shock results in Bright *et al.* 2023). The array has a similar sensitivity and baseline length (and hence angular resolution) to AMI-LA, but has greater frequency flexibility. The ATA can place dual receiver tunings anywhere in the 1–10GHz frequency range, and hence is able to provide radio spectral information more easily than AMI-LA (for which the only option

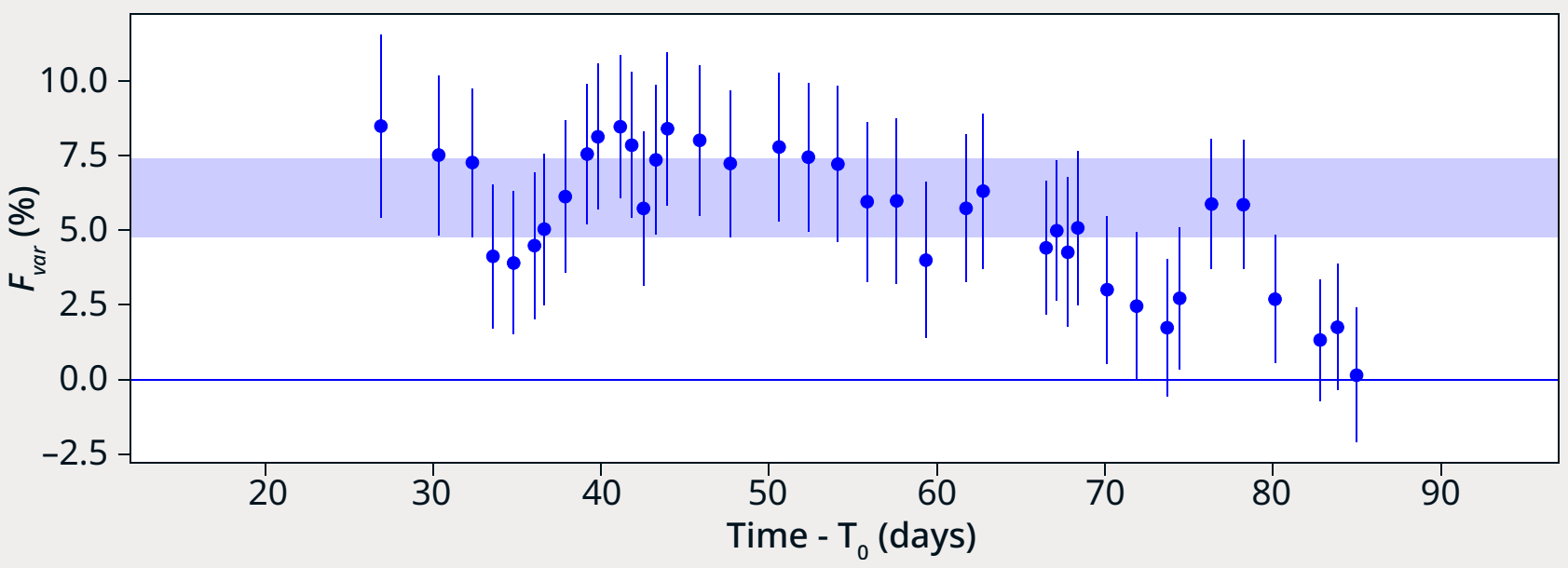

**3** *The fractional daily variability captured by AMI-LA observations in the radio emission of the relativistic TDE AT2022cmc* (Rhodes et al. 2023)*. These measurements were used to constrain the compactness of the sources and provided measurement of the (large) jet Lorentz factor.*

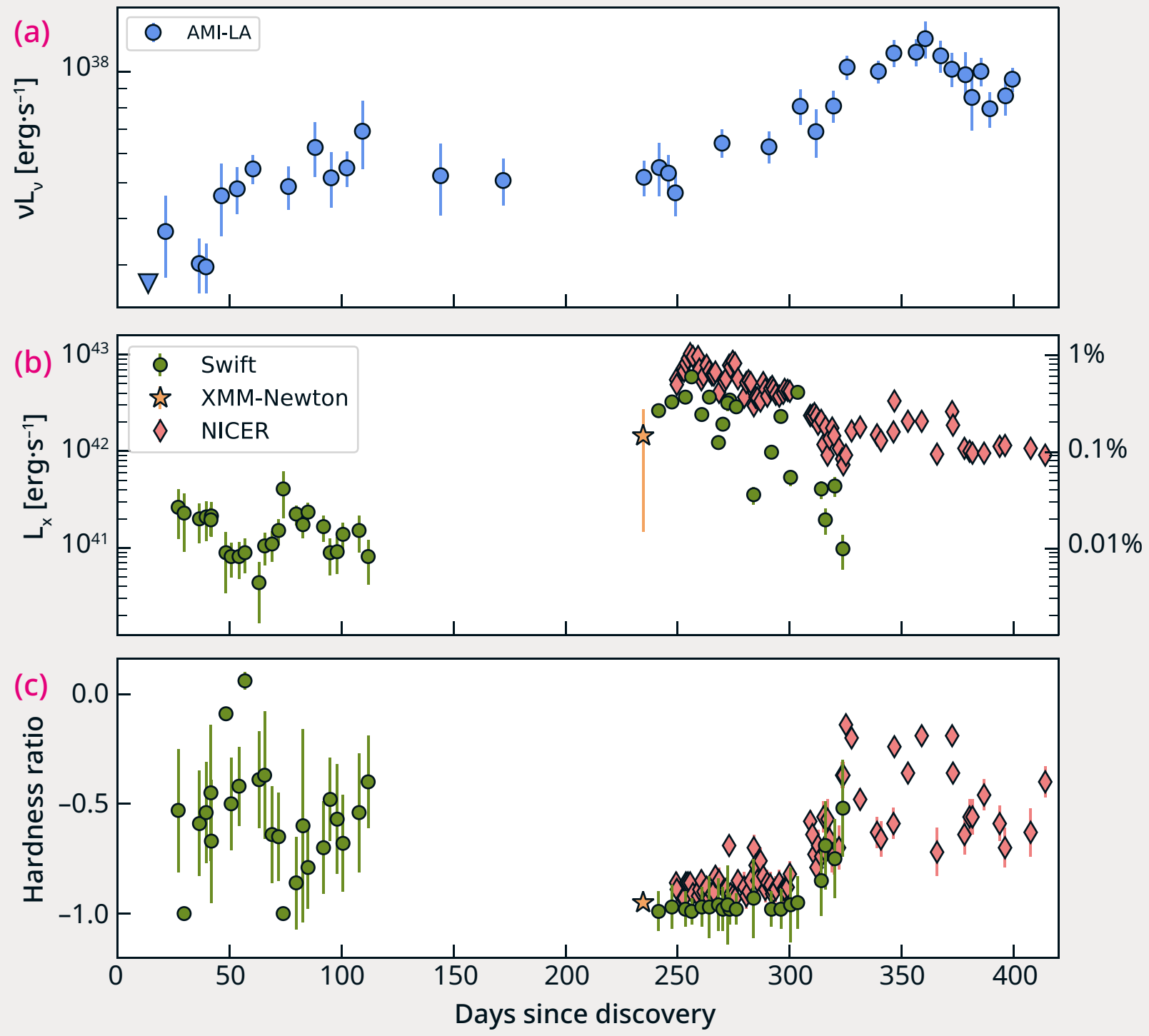

**4 (a)** *The high-cadence AMI-LA observations of the TDE AT2019azh (Sfaradi et al. 2022). The frequent observations revealed a delayed radio flare following the initial radio flare. This flare would have been completely missed in common sparse follow-up data obtained with larger telescopes. Surprisingly, the delayed radio flare followed a delayed X-ray flare* **(b)** *during which the X-ray emission has become completely soft* **(c)**.

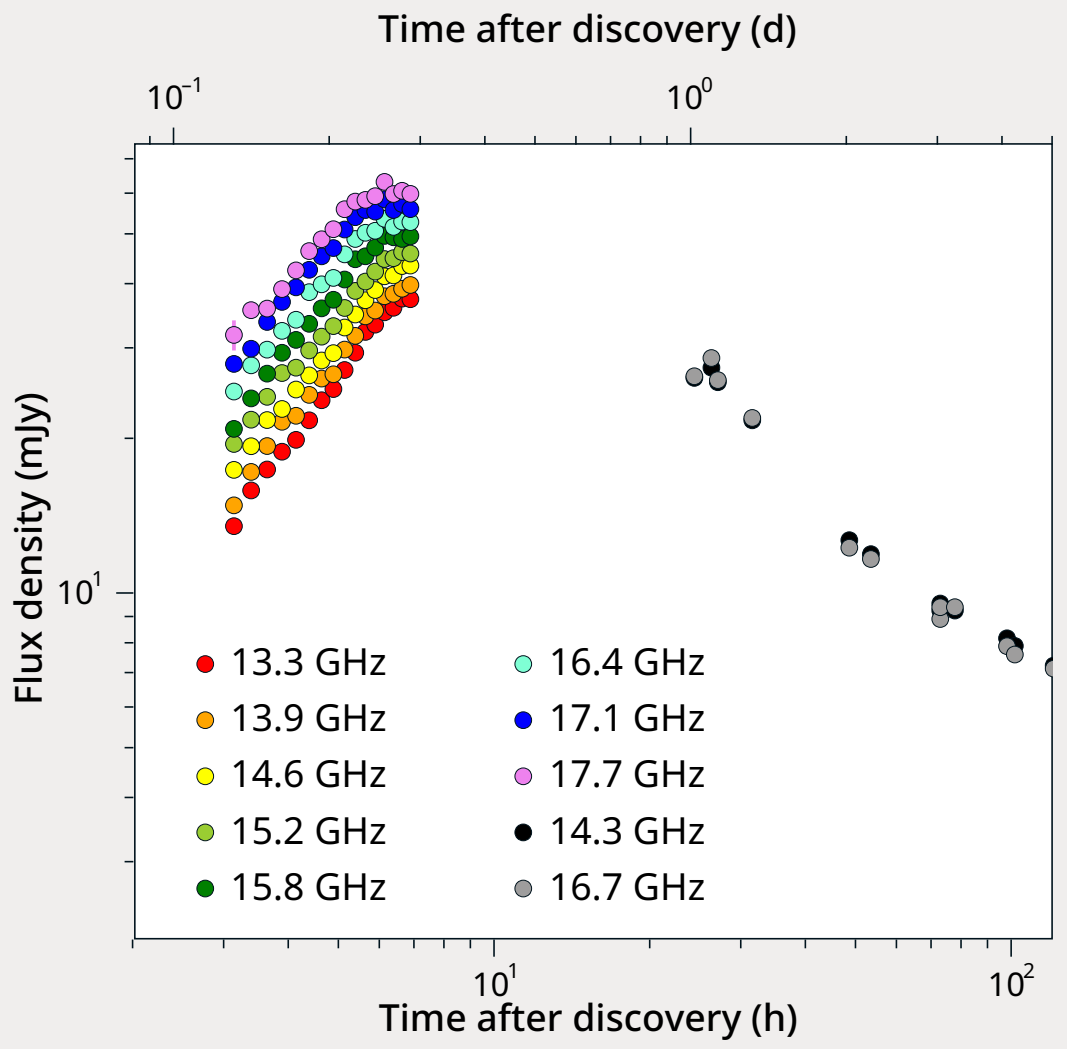

**5** *Very early time high-frequency radio observations of reverse shock in GRB 221009A ('The BOAT') obtained with AMI-LA, a day before any other radio facility observed the source. These observations tested reverse-shock models in an unprecedented way* (Bright et al. 2023).

## A comparison of three small radio arrays: AMI-LA, the ATA and KAT-7

Here we highlight the properties of three small radio arrays, one of which is well established in its radio transients programme (AMI-LA), one of which is just beginning a new journey into this area (ATA), and one of which we think has enormous potential for providing a southern hemisphere facility (KAT–7).

AMI-LA is situated at the University of Cambridge's Lord's Bridge site, at a latitude of +52N. It comprises eight 13-metre dishes operating in the 13–18GHz frequency range, and has baselines in the range 18–110m. The Allen Telescope Array (ATA), operated by the Search for Extraterrestrial Intelligence (SETI) Institute, has recently begun observing radio transients as part of its science programme. The ATA consists of 42 6.1-metre antennas with baselines between 10m and 323m of which 20 antennas are currently recording data. These are equipped with a unique feed design which allows them to be simultaneously sensitive to radiation between 1GHz and 10GHz, of which two independent ~700MHz chunks can be correlated. The recorded bandwidth is only limited by computing power, and can be upgraded in the future. The array has great potential for image plane transient monitoring, especially with its ability to measure instantaneous spectral indices. The ATA is at latitude +41N but its longitudinal separation from the AMI-LA and KAT-7 allow for rapidly evolving transients to be observed near continuously for favourable declinations. With 20 operational antennas the total collecting area of the ATA is approximately half that of the AMI-LA, and when fully operational it will be comparable. The baselines and collecting areas of AMI-LA and ATA are rather similar, and also to those of KAT-7, which has seven 12-metre dishes (i.e. KAT-7 has 75% of the collecting area of AMI-LA), and baselines in the range 26-185m. KAT-7 previously operated in the 1.4GHz band, but the SKA-SA office has indicated that the dishes would be good to at least 5GHz (maybe higher). KAT-7 sits at a latitude of –30S, potentially providing excellent geographic complementarity to AMI-LA and the ATA.

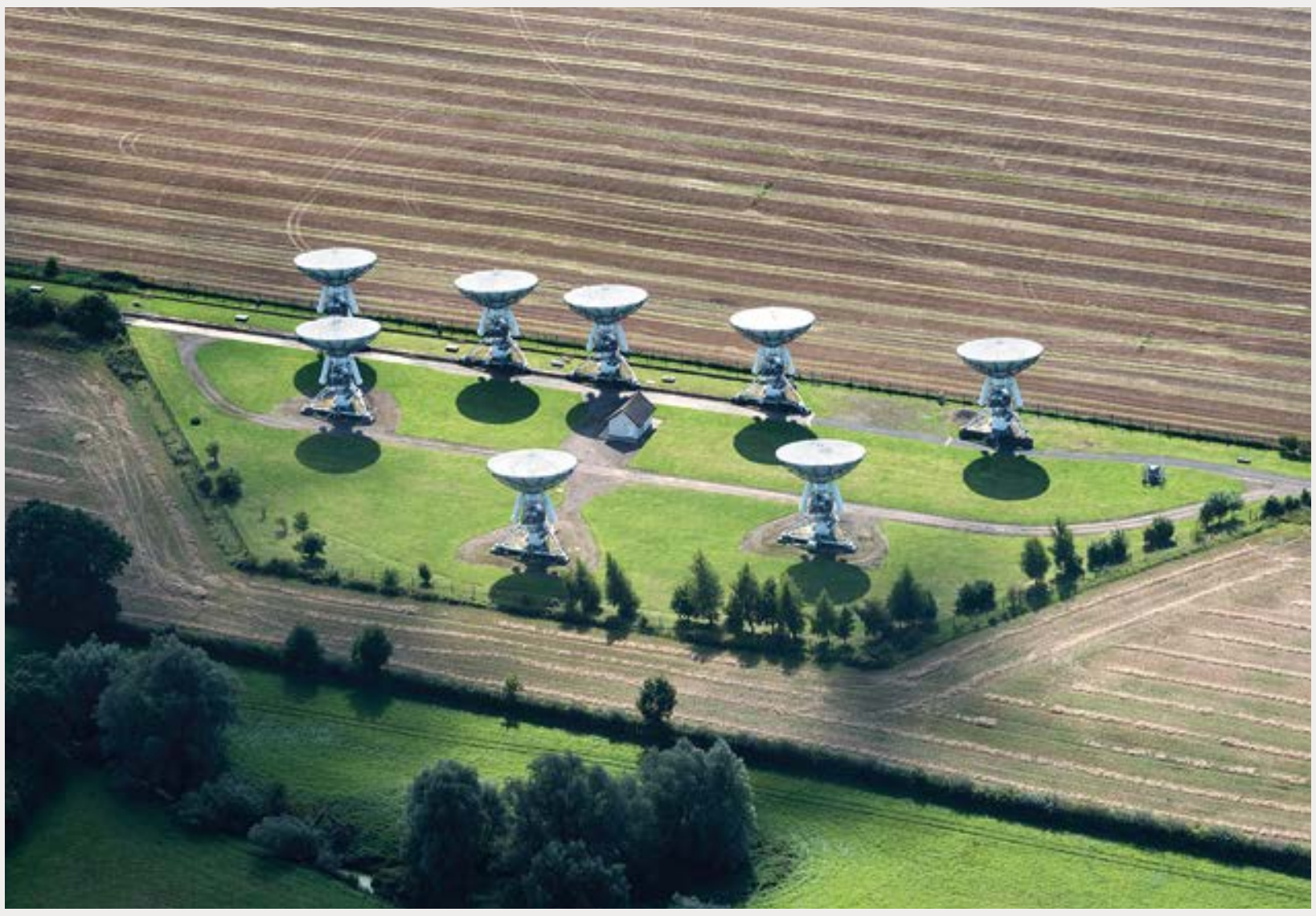

A1 *Aerial view of the AMI-LA array in Cambridge, UK. The longest baseline is 110m. This telescope has for many years performed high-cadence follow-up of a wide range of astrophysical transients and variables. AMI-LA is operated by the University of Cambridge.* (John Fielding Aerial Images)

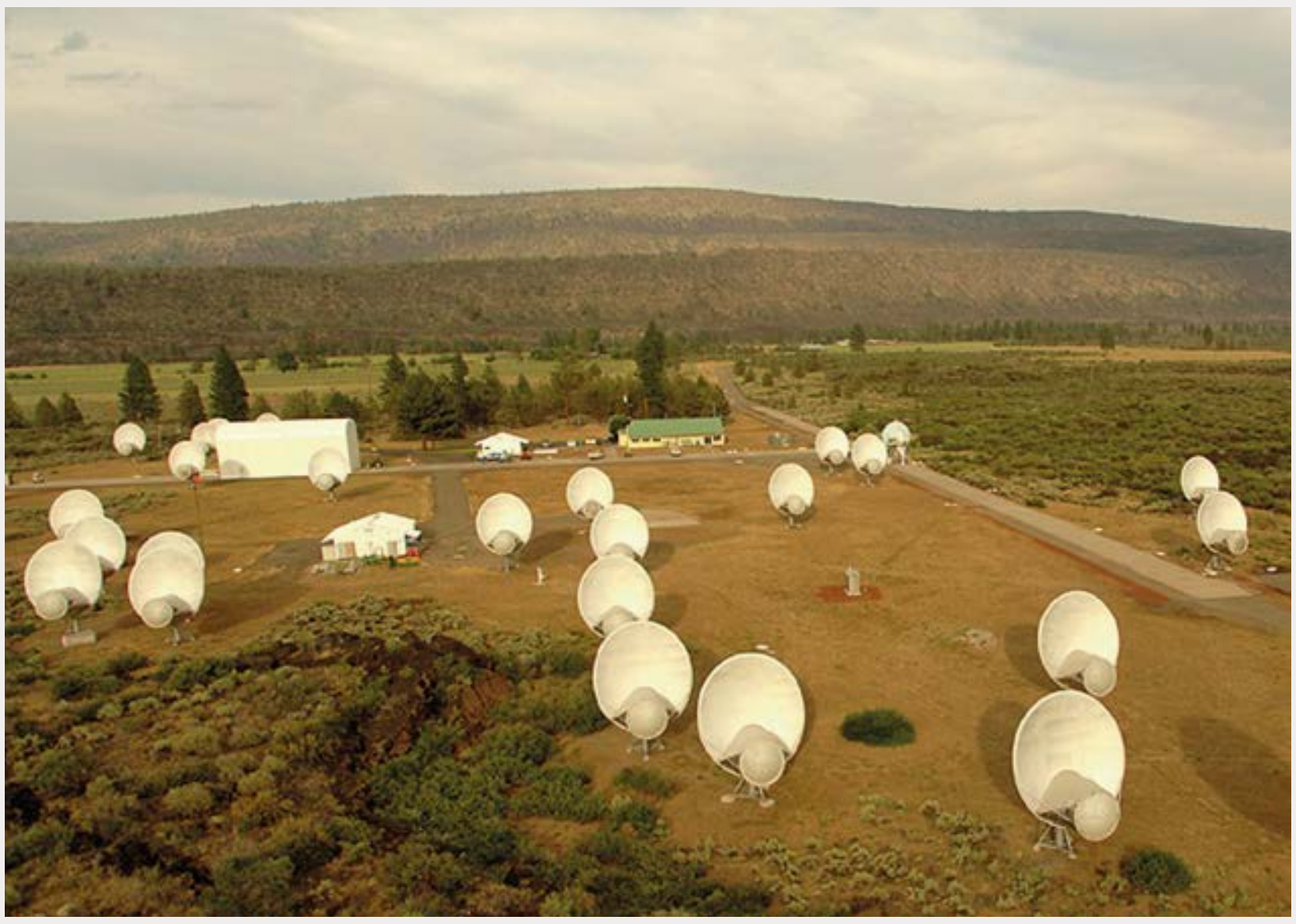

A2 *Aerial view of The ATA located in Hat Creek, California, USA. The longest baseline is 323m. This array has recently recommenced observations of astrophysical transients in parallel with a SETI programme. The ATA is operated by the SETI institute.* (Seth Shostak/SETI Institute)

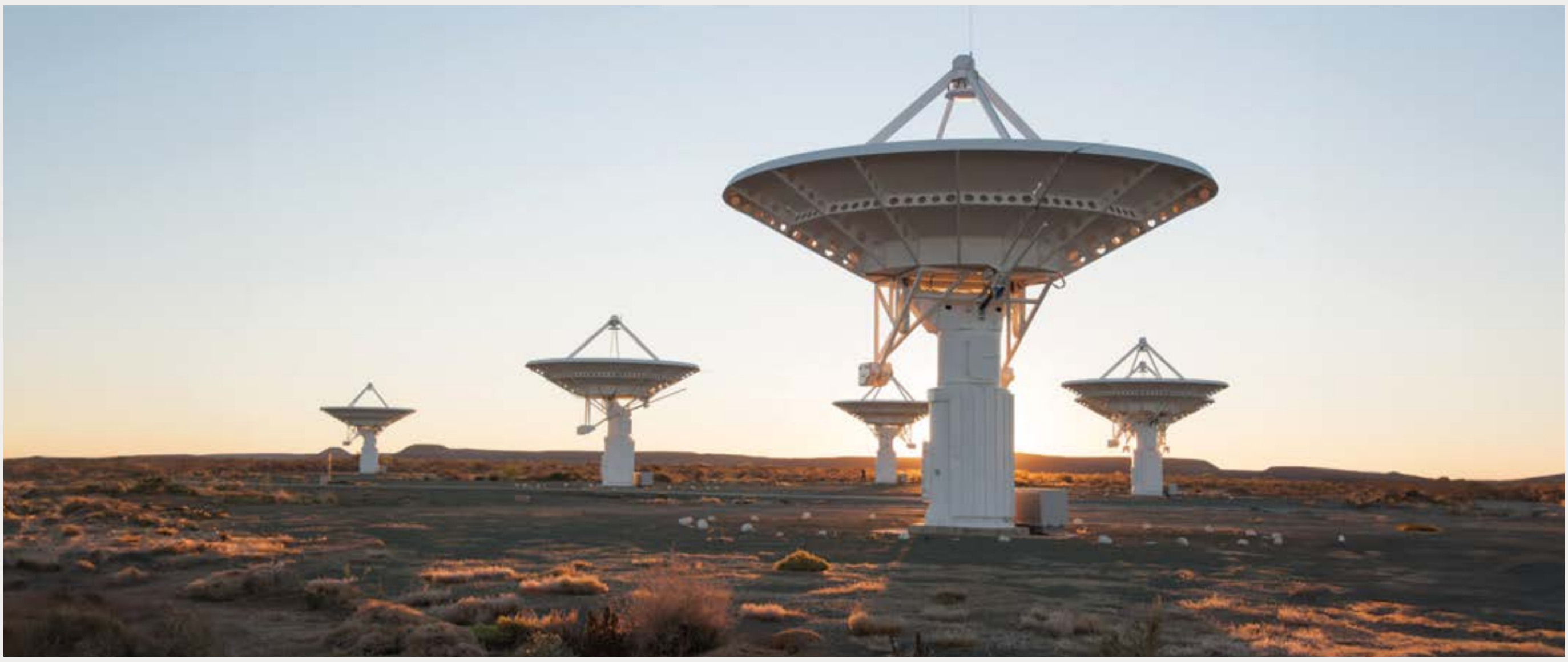

A3 *Antennas of the KAT-7 array located near the MeerKAT (soon to be SKA1-MID) site in the Karoo, South Africa. The longest baseline is 185m. KAT-7 is operated by the South African Radio Astronomy Observatory, SARAO.* (SARAO) is in-band spectral indices for the brightest sources). Even two-point spectral indices can tell us if a radio outburst is optically thick or thin, or whether a radiative cooling break has passed through the band, dramatically altering our interpretation of events.

Both AMI-LA and the ATA are firmly in the northern hemisphere, and we need radio transients coverage in the south. Not only is it the south where the SKA will operate, alongside most of the world's most powerful ground-based facilities at optical and other wavelengths, but it also affords us a view of most of the galactic plane, the Magellanic Clouds and, crucially, the galactic centre.

## Another comparison with optical astronomy

We discussed above how optical astronomy has evolved to a state of a small number of very large facilities supported by a large and growing army of small optical telescopes, often operating robotically. The argument made was that this arose from the need to maintain both wide fields of view and high sensitivity. However, there is another reason, which resonates strongly with our case here: and that, again, is that small facilities with rapid response, high cadence and the ability to monitor for long durations make scientific breakthroughs that the larger optical telescopes will never have the available time to make.

There are notable examples of certain types of transients for which both dedicated long-term monitoring and high-cadence observations played a key role in the discovery.

- PTF11agg (Cenko *et al.* 2013) is a fast optical event discovered in a high-cadence observation of the Beehive cluster. Following discovery, the event faded exponentially with a decay time of five hours. Radio observations of this event indicated the presence of relativistic ejecta. Cenko *et al.* suggested that at odds of 1:100 this event is a regular GRB for which high energy emission was missed (e.g. due to incomplete satellite coverage) or, at more reasonable odds, this is the first example of a dirty fireball (Dermer *et al.* 1999).
- More recently, a new type of explosion has emerged. These are fast blue optical transients (FBOTs; e.g. AT2018cow, Prentice *et al.* 2018), whose optical light curve peaks on a time scale of around two days. They have been discovered thanks to high-cadence optical surveys and are thought to be the case of a rare type of stellar death involving a compact central object.
- Dedicated long-term optical monitoring of transients has also led to new discoveries. iPTF14hls (Arcavi *et al.* 2017b) is an example of a new type of transient, dubbed a 'zombie star' in which a supernova explosion seems to keep on going and being rejuvenated for years.
- Last but not least, a large number of high-cadence observations were key in characterising the fast evolving optical light curve of the first (and so far only) discovered EM counterpart of the neutron star merger gravitational wave event GW170817 (e.g. Arcavi *et al.* 2017a; Cowperthwaite *et al.* 2017; Drout *et al.* 2017; Kasliwal *et al.* 2017, Valenti *et al.* 2017).

## Radio transients facilities in the south

The SKA will be the largest and most powerful radio telescope in the southern hemisphere once it absorbs MeerKAT and begins full operations around 2030 as SKA1-MID (a low-frequency dipole antenna-based array, SKA1-LOW, will be sited in Australia). The science case for SKA is vast and transformational (Braun *et al.* 2015). It will quite rightly be oversubscribed for a long time. Furthermore, observers wishing to fully utilise the incredible raw sensitivity of the array to test, for example, the formation and evolution of radio and star-forming galaxies at the highest redshifts, will not wish to see a subset of the available antennas removed from their array to perform transient monitoring. While, of course, the SKA will observe transients, we think that a huge community service, and a relief for SKA1-MID scheduling, would be provided by a low-operational-cost, high-cadence transient monitoring array in the southern hemisphere. This really is the realisation of the more general arguments made above, and builds upon the two decades of experience we have with AMI-LA and the VLA, and other arrays. The latter is much more powerful, can sub-array, and yet can never and will never match the high-cadence discovery space for transients of AMI-LA. Of course, for very faint cosmological transients like the neutron star merger and LIGO/Virgo gravitational wave burst source GW170817, the VLA is unmatched, and large arrays will always play a central role in transients science. But they will not be able to do it all. If we do not have such an array, then SKA1-MID will provide occasional snapshots of transients, with extremely high signal to noise but little dynamical/temporal context, without which the science will be hamstrung and discoveries missed. A small, high-cadence radio transients array would in fact be the ideal facility for providing triggers for SKA1-MID when sources enter particularly interesting radio states, and so the arrays would be highly complementary.

> *"Without this array we will be left with sparse observations of highly variable phenomena, which will not reveal the temporal variability and multiwavelength connections so vital to understanding astrophysical transients"*

What would such an array look like in terms of collecting area and baseline length? Obviously, the more antennas and longer baselines the better, but these go hand-in-hand with increased cost. The ideal array would be something whose operating costs were a fraction of those of the main facility, i.e. the SKA in the south. Note that baselines long enough to spatially resolve any ejecta from these transients is not required; both AMI-LA and ATA really operate as flux density monitoring facilities. What we have seen is that facilities of the scale of AMI-LA and the ATA can contribute in a major way to transients astrophysics. There will of course be new populations of fainter transients discovered with SKA and other much larger facilities, which a small array will not be able to contribute to.

If we accept these arguments, what are the options for the south? The KAT-7 array was a highly successful demonstrator of South Africa's ability to build a powerful radio interferometer. The first published result from the array was based on observations of the radio transient accreting neutron star Cir X-1, reporting a return to very high levels of radio flaring not seen for two decades (Armstrong *et al.* 2013). The array (like the ATA also) has similar baselines and collecting area to AMI-LA (see 'A comparison of three small radio arrays'), and the dish surfaces are good for observations up to 10GHz or better. It has not been used for several years but located very close to MeerKAT (soon to be the SKA1-MID core) and could likely be recommissioned as a transients monitoring array. Other possibilities in the south include eventually repurposing ASKAP or ATCA in Australia or, of course, simply building a new small array (perhaps as part of SKA development).

## On commensality and single millimetre dishes

Commensal radio transients, those found serendipitously in observations made with other targets, are not a focus of this discussion and in fact the majority may well be discovered in the deeper observations made by e.g. SKA (or ngVLA in the north), although again discovery space potential depends upon multiple revisits to the same part of the sky. Follow-up of the most interesting of these transients, if bright enough, could well be part of the programme of the proposed array.

A contribution to transient follow up and monitoring can also be made with single dishes operating in the millimetre range. The reason this single dish approach does not work at lower radio frequencies is that unless the dish is very large, the field will contain much more flux density than that associated with the transient, and without interferometric imaging from an array it is essentially impossible to identify the flux density from any transient. A millimetre dish has some advantages however: the astronomical field of view is much smaller for a given dish size and hence contains many fewer confusing sources, most radio sources on the sky have optically thin synchrotron spectra ($S_\nu \propto \nu^\alpha$, with $\alpha$ typically around $-0.6$), and flaring synchrotron sources, which are initially optically thick, peak earlier and brighter at higher frequencies. For example, the optically thick GRB reverse shock from GRB2210009A caught by AMI-LA in the 16GHz band (and shown in figure 2) would have peaked at around 0.4Jy at 230GHz around 35 minutes post-burst. Since many (but certainly not all) of the most interesting radio transients produce multiple flaring events which are likely to also be associated with strong flaring, millimetre transients monitoring is a very interesting and complementary possibility. When transients become optically thin and faint, at later times in outbursts, the high frequencies become a disadvantage however, so rapid response needs to be folded into a millimetre transients programme.

## Options in the south?

One facility with a good deal of exciting potential is the Africa Millimetre Telescope, a 15m dish, equipped with receivers up to 230GHz, to be built in Namibia in the second half of this decade. Although its main purpose is to provide a valuable south-eastern baseline for the Event Horizon Telescope, when not operating in this role it will perform an extensive high-cadence and rapid-response transient monitoring programme. Millimetre arrays such as ALMA, NOEMA etc. can of course contribute in a very major way to transients science, but these powerful and flexible facilities are oversubscribed in a similar way to VLA/SKA and will probably never be able to dedicate large fractions of their time to transients.

We believe that the case for having a small dedicated radio transients array operating in the south is overwhelming. While we do not attempt to cost any potential options, it is clear that the operational costs of such an array would be a tiny fraction of those associated with running SKA or ngVLA. Without this array we will be left with sparse observations of highly variable phenomena (albeit ones of incredibly high quality) which will not reveal the temporal variability and multiwavelength connections so vital to understanding astrophysical transients. In return, this array would remove pressure from the SKA-scale arrays to perform such high-cadence observations. We note that single millimetre dishes can also contribute to this science, usually in the early phases of a transient, if operated with a rapid-response policy. ●

AUTHORS

**Rob Fender** (rob.fender@physics.ox.ac.uk) is head of astrophysics at the University of Oxford. He has a long standing interest in finding and understanding astrophysical transients, particularly in the radio band. He was the recipient of the 2020 Herschel medal of the Royal Astronomical Society.

**Assaf Horesh** (assafh@mail.huji.ac.il) is an associate professor at the Hebrew University of Jerusalem, Israel. His research focus is time domain astronomy, and includes the exploration of various types of transient phenomena including: supernovae, tidal disruption events, gamma ray bursts, gravitational wave EM counterparts, and new classes of transients.

**Phil Charles** (p.a.charles@soton.ac.uk) is an emeritus professor at the University of Southampton, a former director of SAAO, and current UK SALT Consortium director on the SALT Board. He is active in the SALT Large Program on Transients, using a wide variety of space and ground-based facilities.

**Patrick Woudt** (Patrick.Woudt@uct.ac.za) is the interim dean of the Faculty of Science at the University of Cape Town, and the co-PI of the ThunderKAT large survey project on MeerKAT. His research interests cover accretion processes and accretion-related outflows in cataclysmic variables and related systems.

**James Miller-Jones** (James.Miller-Jones@curtin.edu.au) is a radio astronomer with expertise in high angular resolution observations and scientific interests in accreting compact objects and their jets. He is currently the science director for the Curtin University node of the International Centre for Radio Astronomy Research.

**Joe Bright** (joe.bright@physics.ox.ac.uk) is a researcher in Radio Astronomy at the University of Oxford, UK, and Breakthrough Listen. He is interested in radio emission from transient and variables sources, particularly X-ray binaries, gamma-ray bursts, and exotic supernovae.